\documentclass{article}
\usepackage[utf8]{inputenc}
\usepackage[a4paper, margin=20mm]{geometry}
\usepackage{amsmath,amssymb,amsfonts}
\usepackage{placeins}
\usepackage{authblk}
\usepackage{graphicx,color}
\usepackage{xcolor}
\usepackage{listings}
\usepackage{siunitx}
\usepackage{hhline}
\usepackage{hyperref}
\usepackage{wrapfig}
\usepackage{algorithmic}
\usepackage{textcomp}
\usepackage{cite}
\usepackage{subcaption}

\title{A merging interaction model explains human drivers' behaviour from input signals to decisions}

\author{Olger Siebinga, Arkady Zgonnikov, and David Abbink}
\date{December 2023}

\begin{document}

\maketitle

\begin{abstract}
One of the bottlenecks of automated driving technologies is safe and socially acceptable interactions with human-driven vehicles, for example during merging. Driver models that provide accurate predictions of joint and individual driver behaviour of high-level decisions, safety margins, and low-level control inputs are required to improve the interactive capabilities of automated driving. Existing driver models typically focus on one of these aspects. Unified models capturing all aspects are missing which hinders understanding of the principles that govern human traffic interactions. This in turn limits the ability of automated vehicles to resolve merging interactions. Here, we present a communication-enabled interaction model based on risk perception with the potential to capture merging interactions on all three levels. Our model accurately describes human behaviour in a simplified merging scenario, addressing both individual actions (such as velocity adjustments) and joint actions (such as the order of merging). Contrary to other interaction models, our model does not assume humans are rational and explicitly accounts for communication between drivers. Our results demonstrate that communication and risk-based decision-making explain observed human interactions on multiple levels. This explanation improves our understanding of the underlying mechanisms of human traffic interactions and poses a step towards interaction-aware automated driving. 
\end{abstract}

\section{Introduction}
\begin{figure}[t!]
\centering
\includegraphics[width=\linewidth]{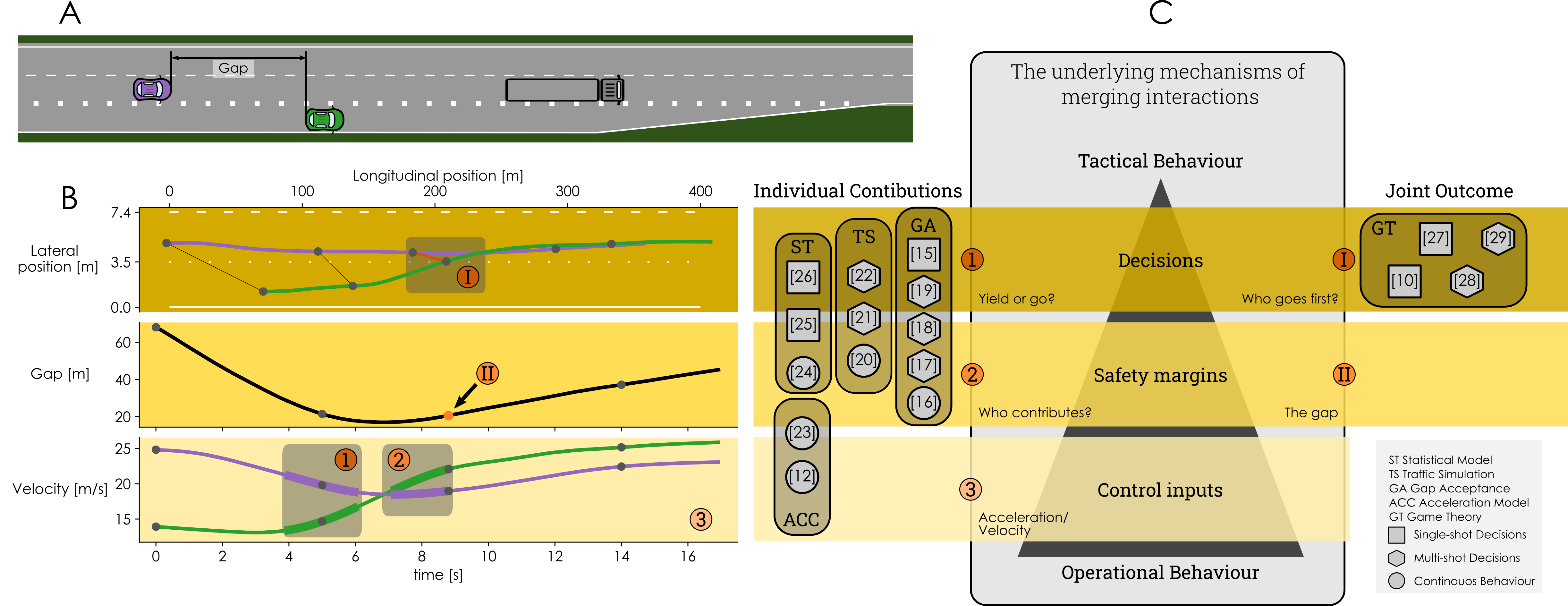}
\caption{Models of highway interactions and the aspects of interaction they describe in a merging scenario. \textbf{A}: a typical interactive merging scenario taken from the HighD dataset~\cite{Krajewski2018} (dataset 60, vehicles 458 and 468). In this scenario, the driver of the green vehicle wants to merge onto the highway. This vehicle has a position advantage but a significantly lower velocity compared to the purple vehicle. \textbf{B}: the vehicles' position traces, the gap between the vehicles, and the individual velocity traces. In this example, the joint high-level decision is indicated at I: Green merges ahead of Purple (i.e., Green goes first). Both vehicles individually contribute to this decision by accelerating and decelerating respectively (at 1). After the decision has been made, Green keeps accelerating and thereby individually contributes to maintaining a safety margin while purple stops decelerating (2). The gap between the vehicles, when Green crosses the lane marker (II), denotes the joint safety margin. Finally, the underlying characteristics of the individual vehicle control inputs are depicted here as the total velocity traces (3). We evaluate velocity traces instead of raw accelerations because they are easier to perceive for other drivers and provide more insight into the trend of the driver's actions since they are less noisy. These individual and joint perspectives on the three levels of behaviour are also indicated in panel C. \textbf{C}: the three levels of behaviour in between Michon's \textit{operational} and \textit{tactical} behaviour~\cite{Michon1985}. It also shows five modelling strategies for merging interactions, each with examples from the literature. The icons indicate if the models describe a single decision at the start of an interaction (one-shot), repeated decisions (multi-shot), or continuous behaviour. Every modelling strategy captures part of the overall interactive behaviour, but none covers all five aspects. We postulate that a model capturing all three levels of individual and joint behaviour simultaneously is likely to have captured the underlying mechanisms of merging behaviour.}
\label{fig:literature_overview}
\end{figure}

Automated driving holds many potential benefits for society~\cite{Harper2016, Clements2017, Pettigrew2017}, but, safe and efficient interactions between Automated Vehicles (AVs) and human-driven vehicles remain an open problem~\cite{Brown2023b}. Such interactions frequently occur in everyday traffic: at intersections, on roundabouts, and on highways. This work will focus on merging interactions on highways as they are especially intricate due to the high speeds and multiple available options to resolve a conflict (Figure~\ref{fig:literature_overview}-A). 

A potential solution to handling such interactions in AVs is through interaction-aware controllers (e.g.,~\cite{Sadigh2018, Schwarting2019}). These controllers assume that human drivers unilaterally respond to the AV's behaviour and use a model to predict these responses~\cite{SiebingaIACValidation}. However, real-world merging interactions are inherently \textit{reciprocal}~\cite{SiebingaCEITheory}: a driver does not only respond to another driver but also influences their behaviour through implicit (or even explicit) communication~\cite{Brown2023a, Brown2023b}. Individual control inputs and decisions of two or more drivers in a merging situation lead to a \textit{joint interaction} outcome on multiple levels (Figure~\ref{fig:literature_overview}-C): high-level decision-making (negotiating who goes first), acceptable safety margins, and required individual control inputs. This makes real-world merging behaviour complex to understand and model, both from an individual and joint perspective (for a real-world example, see Figure~\ref{fig:literature_overview}-B). Interaction-aware AVs should use a model that captures this complexity, which is currently lacking. 

Previous work has shown that in merging interactions, \textit{high-level individual decisions} are made to yield or to go~\cite{Kita1999, Brown2023b}, which lead to universal joint outcomes in terms of who merges first based on the kinematics of a merging scenario~\cite{SiebingaEmpirical}. The \textit{safety margins} (e.g., gaps between two vehicles) are the result of joint behaviour~\cite{Treiber2000}, but at the same time, these gaps are used by individual drivers to communicate their intent~\cite{Brown2023a}. \textit{Low-level control inputs} (i.e., acceleration, velocity, and position) are used by individual drivers to communicate~\cite{Kauffmann2018, Brown2023a}, thereby playing an essential role in both the \textit{outcome} of the interaction and the \textit{human perception} of other vehicles' behaviour. These interrelated aspects of individual and joint behaviours are not well understood, and a driver model capturing all aspects is missing.

Our work builds on related work in merging and lane-changing models (merging is often considered a special type of lane change~\cite{Zhang2020}), in which we identified five classes: 1) gap acceptance; 2) traffic simulation; 3) statistical; 4) acceleration; and 5) game theoretic models (Figure~\ref{fig:literature_overview}-C). Gap acceptance models (e.g.~\cite{Kondyli2011, Laval2008, Choudhury2007, Michaels1989, Ahmed1999}) describe the decisions made by the individual merging drivers by evaluating available gaps (safety margins) against a personal minimal acceptable gap size. Traffic simulation models often rely on the same gap acceptance theory for making high-level decisions~\cite{Kesting2007, Yang1996, Hidas2002}, and are complemented with acceleration models (e.g., the intelligent driver model (IDM)~\cite{Treiber2000}) to include the control behaviour before and after the merging decision. These acceleration models describe individual accelerations and do not include interactive or communicative behaviour. We found one acceleration-based model that describes individual drivers' control inputs and safety margins during interactions~\cite{Wan2014}. Statistical models provide a probability that a certain vehicle will merge or change lanes based on naturalistic traffic data. Some include desired safety margins~\cite{Daamen2010, Marczak2013}, while others do not~\cite{Dong2018}. Finally, game-theoretic models describe the high-level outcome and decision-making of multiple drivers in a single model (e.g.,~\cite{Kita1999, liu2007, coskun2019, Li2018}, see~\cite{Ji2020a} for a review). Game theoretic models assume humans to be rational utility-maximizing agents that do not communicate. However, it is known that these assumptions do not hold for merging drivers~\cite{SiebingaCEITheory}.

To achieve predictable, legible~\cite{Dragan2013}, acceptable, and safe automated behaviour, we need to provide interaction-aware AVs with a driver model that covers all three levels. The decision level is important because if automated driving violates the underlying behavioural norms of human drivers on this level (e.g., it claims the right of way) its behaviour will be unacceptable to passengers and other drivers~\cite{Brown2023b}. Automated behaviour should adhere to acceptable safety margins (on a joint level) and understand how individual drivers keep these margins. This way, AVs will show behaviour that is not just safe but is also perceived as safe. Finally, understanding the subtleties of gaps, positions, and velocities can help AVs understand the communication from other drivers and act accordingly. Since merging is a reciprocal interaction, such a model should capture the joint behaviour, not just that of a single driver responding to their environment~\cite{SiebingaCEITheory}. Since merging is a reciprocal interaction, such a model should capture the joint behaviour, not just that of a single driver responding to their environment~\cite{SiebingaCEITheory}. 

Besides direct applications in interaction-aware AVs, a complete model of merging interactions could also prove to be a valuable step towards theories and a better fundamental understanding of human interactive capacities and behaviour in general~\cite{Guest2021}. A joint driver model could be used to understand and investigate how drivers perceive the behaviour of others, how they communicate, and how they negotiate a safe solution in general traffic interactions. This fundamental understanding is needed to design automation that can interact in a natural manner. In a review of automated interactive traffic behaviour, Brown et al. stated: "Designing systems that can understand and react to such [implicit traffic] communication will rely upon developing an understanding of that communication beyond statistical regularity"~\cite{Brown2023b}. Black-box trajectory prediction models (e.g.,~\cite{Salzmann2020, Brito2020, Meszaros2023}) do not provide insight into the underlying mechanisms of merging interactions and will thus not suffice for this purpose. If a model of merging interactions succeeds in capturing the underlying mechanisms of merging interactions, which is likely if it captures behaviour across multiple levels, it could generalise to other interactive traffic scenarios, helping us gain insight into the fundamentals of interactive human driving behaviour.

The main contribution of this manuscript is a novel computational model for a simplified merging scenario with human drivers, based on the Communication-Enabled Interaction (CEI) framework~\cite{SiebingaCEITheory}. The model assumes that drivers have a deterministic plan for the near future and form a probabilistic belief about another driver's intentions based on implicit communication. The plan and belief result in a perception of risk. If this risk exceeds a personal threshold, a driver updates their plan to get the risk under control. We validate our model on empirical data collected in a top-down view driving simulator with pairs of drivers~\cite{SiebingaEmpirical, SiebingaExperiment}. Our model accurately describes the (qualitative and quantitative) control input characteristics, safety margins and high-level decisions (i.e., who goes first?) of human drivers. It captures differences in individual contributions of drivers and joint behaviour. Finally, our model does not assume human rationality and explicitly incorporates communication between drivers as one of the fundamental aspects of interactions, making it the first merging interaction model to avoid these common game-theoretic assumptions.

\section{Results}
\begin{figure}[ht!]
    \centering
    \includegraphics[width=\textwidth]{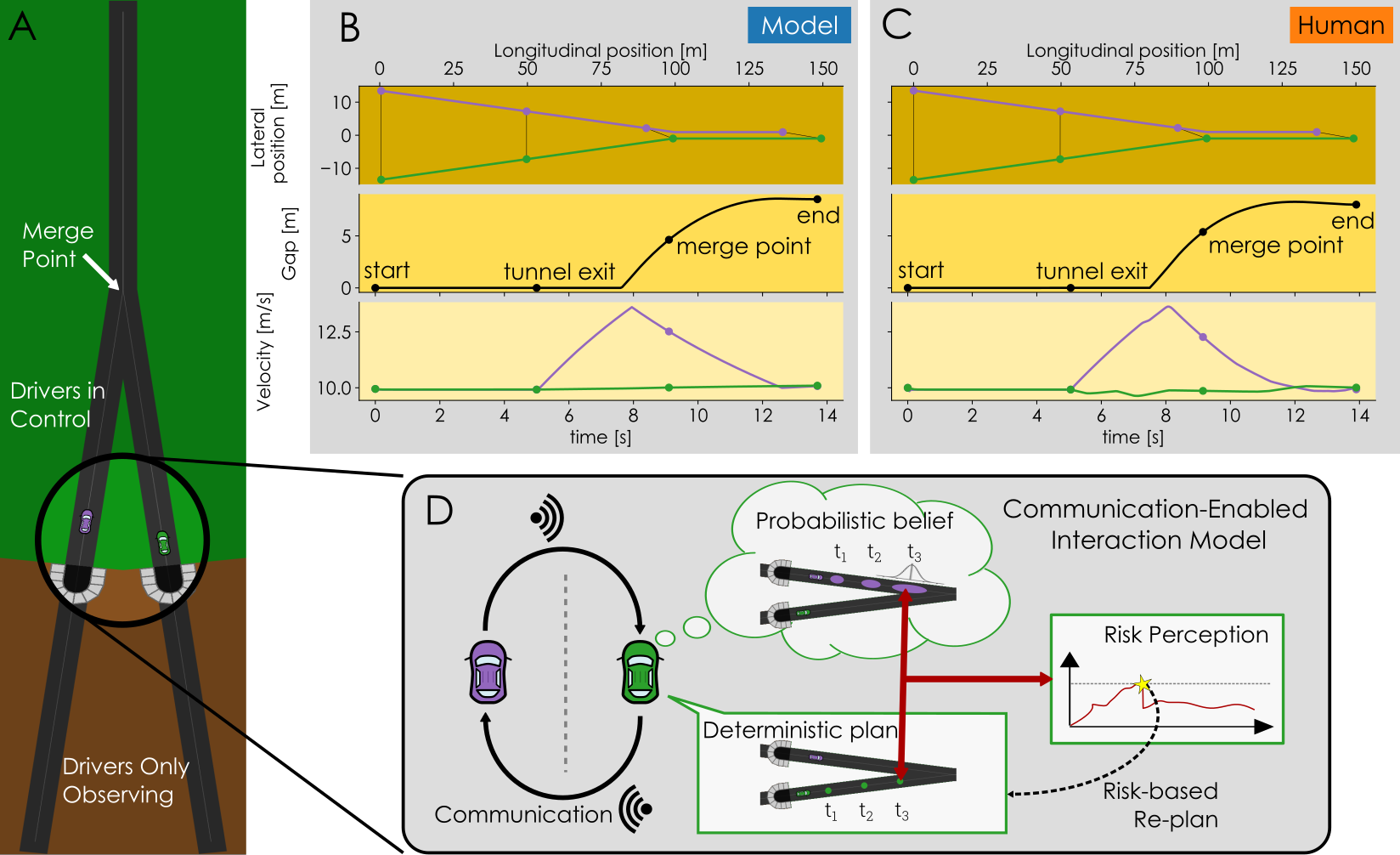}
    \caption{\textbf{A}: the simplified merging scenario used both in model simulations and the experiment in a coupled top-down-view driving simulator~\cite{SiebingaEmpirical}. Two vehicles start on different roads, which they follow to a single merge point of those roads. They start in a tunnel where the driver (i.e., either human participants or the model) can observe both vehicles and their initial velocities but have no control yet. After exiting the tunnel, the drivers can now control the vehicles' acceleration (no steering control is needed or available). Beyond the merge point is a short road where the vehicles follow each other. \textbf{B and C}: typical examples of human and model interactions in this scenario (participant pair 3, the model behaviour resulted from a fit on the human behaviour across all conditions and all trials; the model trial shown here was not fitted specifically to this human trial). \textbf{D}: The Communication-Enabled-Interaction (CEI) framework. For clarity, the panel only visualises the three model components for the green vehicle, but the model is symmetrical, so the purple vehicle has the same three components. Each driver has a \textit{deterministic plan} for their own behaviour and a \textit{probabilistic belief} of the positions of the other driver in the near future. Combined, the plan and belief result in a continuous perception of \textit{risk}. If this perceived risk exceeds a \textit{risk threshold} the driver alters their plan to return the risk under the threshold. Each driver \textit{communicates} their plan (intention) implicitly (e.g., through vehicle motion) to the other driver, who bases their belief on the received communication. Thus, this communication links one driver's plan to the belief of the other driver.}
    \label{fig:scenario}
\end{figure}

\paragraph{Simplified merging scenario and experiment}
To study and model merging interactions between two human drivers, we used data previously gathered in an experiment using a simplified merging scenario (Figure~\ref{fig:scenario}-A) in a coupled, top-down view driving simulator~\cite{SiebingaExperiment, SiebingaEmpirical}. The scenario simplifies merging by simulating two roads (or lanes) that merge into one at a single point. The vehicles start in a  tunnel where the drivers can only observe both vehicles travelling at constant velocity to facilitate velocity perception before the interaction. Once both vehicles have exited the tunnel, the drivers gain control over the accelerations of their vehicles (steering is not possible) to resolve a merging conflict. The drivers (9 pairs of participants) were instructed to maintain their initial velocity yet prevent a collision.
The experiment used 11 different experimental conditions that would end in a collision if the vehicles kept their initial velocity (10 repetitions per condition). A condition consisted of a combination of initial relative velocity ($-0.8, 0.0,$ or $0.8~m/s$) and the projected headway at the merge point if both vehicles would continue their initial velocity ($-4, -2, 0, 2,$ or $4~m$). The names of the conditions denote \textit{projected headway}\_\textit{relative velocity} (e.g., 4\_-8). Positive numbers indicate an advantage for the left driver. We refer to an individual driver as either the left or right driver throughout all trials, based on their physical location during the experiment. The scenario was completely symmetrical (i.e., there was no right of way). Note that from the drivers' view, they randomly perceived approaching the merge point from the left or right side of the track -- to account for potential biases due to traffic rules. They were seated in the same room but could not see each other or communicate in any other way than via vehicle motion. For a more extensive description of the experimental protocol, see~\cite{SiebingaExperiment, SiebingaEmpirical}. Figure~\ref{fig:scenario}-C shows a typical trial outcome with human participants.

\paragraph{Model and Communication-Enabled Interaction (CEI) framework} 
We created a novel model based on the Communication-Enabled Interaction (CEI) framework in~\cite{SiebingaCEITheory} to describe the joint behaviour of a pair of merging drivers. At the core of the CEI framework lies the idea that drivers communicate their plans (intentions) to others using implicit or explicit communication. Empirical evidence has shown that this kind of communication plays an important role in traffic interactions (e.g.~\cite{Lee2021, Brown2023a}), which to date has been absent in interactive driving models. We assume that drivers form a probabilistic belief about the other driver's future movements (intent) based on this communication. Combined with their own deterministic plan, this belief underlies a driver's perceived risk. We assume that if the risk exceeds the driver's individual risk threshold, they will unilaterally alter their plan to get the risk under control. The CEI framework~\cite{SiebingaCEITheory} describes this overall structure consisting of four modules: communication, plan, belief, and risk perception. The model proposed here instantiates the CEI framework by implementing these four modules. 

The methods section provides a full specification of all modules of our model (plan, communication, belief, and risk), yet we give a summary here to help interpret the results. 
In our model (Figure~\ref{fig:scenario}-D), drivers plan a deterministic trajectory (i.e., a set of waypoints) by optimising comfort and speed over a time horizon. They communicate this plan implicitly through vehicle kinematics (current position, velocity, and acceleration). Drivers' velocity perception is assumed to be noisy. The drivers' belief about the actions of the other vehicle is represented as a set of probability distributions for the other vehicle's positions at specific times in the future. The recent behaviour of the other vehicle influences the variability in the belief (i.e., inconsistent behaviour increases the variance). The perceived risk is calculated by evaluating the probability that the positions of the ego and the other vehicle (plan and belief) overlap (i.e., the probability of a collision). 

We assume every driver has two dynamic risk thresholds: an upper threshold and a lower threshold. The plan is updated either when a) the upper threshold is exceeded (to prevent a collision) or when b) the perceived risk stays below the lower threshold for a certain amount of time $\tau$ (to revert to "normal" behaviour when the conflict is resolved). Both thresholds are dynamically adjusted by an incentive function reflecting traffic rules and customs. The rationale behind this is that two drivers perceive the same amount of risk, but traffic rules and customs provide a higher incentive for one of them to act. For example, the following vehicle is usually responsible for preventing collisions in a car-following scenario. 

\subsection*{Model simulations}
The model uses 10 parameters designed to reflect the scenario which were equal for all simulations. We used a grid search to find individual risk threshold parameters to describe the nine pairs (18 drivers) from the experiment~\cite{SiebingaEmpirical} (i.e., one upper and lower threshold per driver is used across conditions). The incentive functions are the same for all drivers and were fitted to all experimental data using linear regression. We simulated the same number of trials as in the experiment: 9 participant pairs, 11 conditions, and 10 repetitions of each condition per pair (990 total trials). The model simulations run faster than real-time with an average run time of $2.6~s$ (Intel Xeon E5 quad-core) for an average real-time duration of $14.2~s$. An example of a simulated trial for participant pair 3 can be found in Figure~\ref{fig:scenario}-B. 

In the remainder of this section, we will evaluate the model behaviour on the three behavioural levels presented in Figure~\ref{fig:literature_overview}-C, both for individual and joint behaviour. Some trials ended in a collision; these are excluded from the results because they represent edge cases. Collisions happened infrequently and in all conditions for human drivers (28/990) and model simulations (29/990). The online supplementary materials contain more details on how the collisions were distributed over conditions. 

\begin{figure}[hp!]
    \centering
    \includegraphics[width=\textwidth]{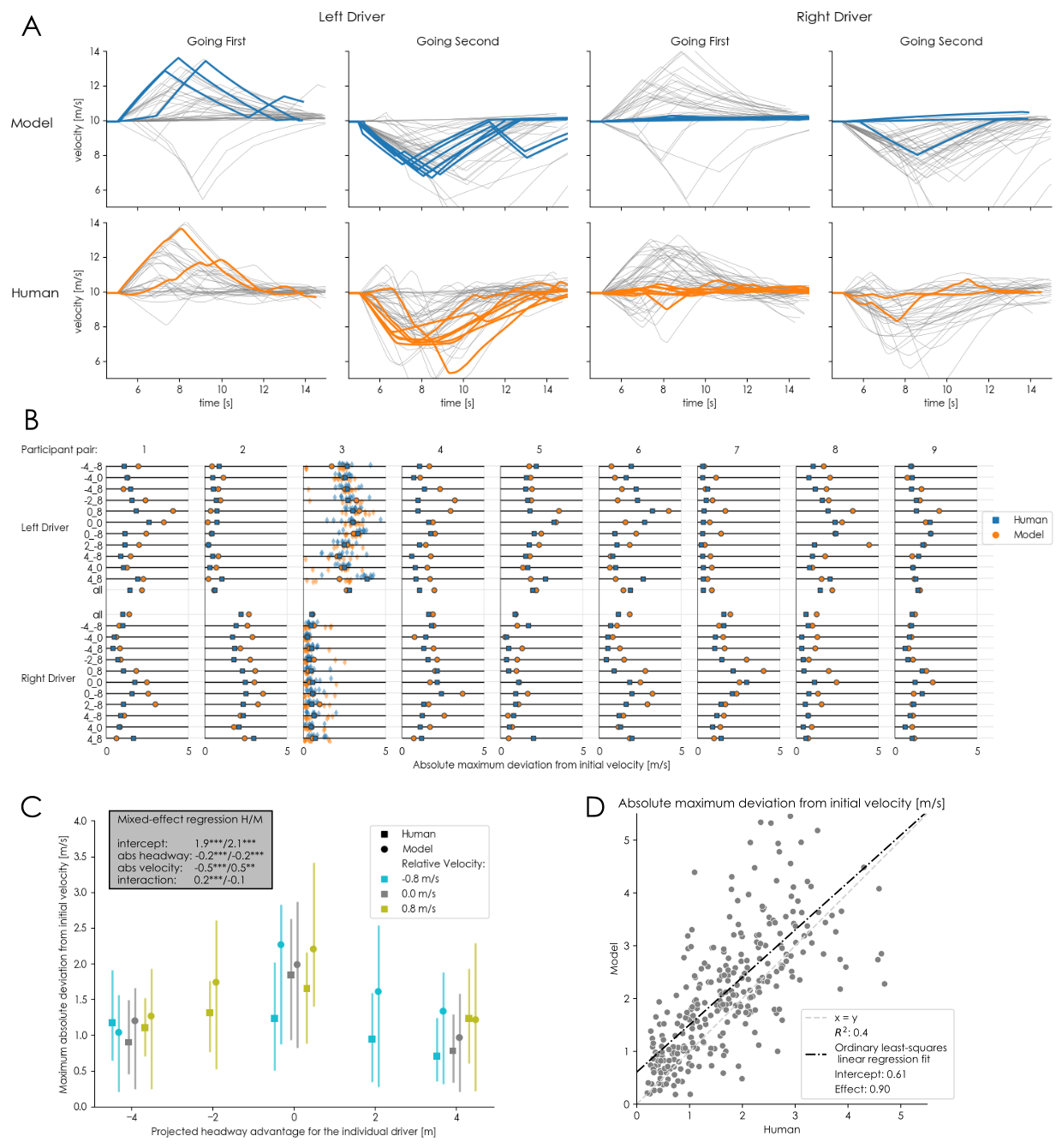}
    \caption{\footnotesize An overview of a comparison of the \textbf{control inputs} performed by 9 pairs of human drivers, and by the model fit to capture these pair's merging behaviour. Trials that ended in a collision were excluded here. \textbf{A}: all velocity traces for the model (top row) and human drivers (bottom row) in condition 0\_0 (i.e., no initial velocity or position difference between the vehicles), for each driver (left/right) in the pair, divided into the high-level outcome of the trial (i.e., which driver merged first). To facilitate the comparison for a single pair, the velocities of pair 3 are highlighted. \textbf{B}: mean absolute maximum deviation from the initial velocity for the left and right driver in all participant pairs for all experimental conditions. We use the absolute deviation to compare conditions because a single condition can contain both outcomes within one pair (left going first and right going first, e.g., see panel A). Thus a single driver could accelerate in some trials and decelerate in others within one condition. The minimum and maximum deviations will be studied separately in section~\ref{sec:decisions} as individual contributions to the decision. For pair 3, all underlying data are visualised (10 trials per condition). One outlier ($9.5~m/s$, model trial, condition 4\_0) is not shown. \textbf{C}: mean absolute maximum deviation from the initial velocity, aggregated over all drivers for different relative velocities and projected headways. The error bars represent interquartile ranges. The headway and velocity values in panel C are shown from the perspective of the individual driver. This means that a projected headway of $4~m$ represents an advantage from the driver's perspective, while $-4~m$ represents a disadvantage. The inset indicates coefficients of mixed-effects linear regression models predicting the mean absolute maximum deviation from initial velocity as a function of projected headway and relative velocity ($*: p<=0.05, **: p<=0.01, ***: p<=0.001$). Full results of the statistical analyses are available in the supplementary materials. \textbf{D}: the relationship between human and model behaviour for all participant pairs and all conditions (i.e., all points from panel B).}
    \label{fig:inputs}
\end{figure}

\subsubsection*{Characteristics and magnitude of control inputs}
Empirical evidence showed that human drivers use intermittent piece-wise constant acceleration control to solve merging conflicts in the simplified merging scenario~\cite{SiebingaEmpirical}. This type of control results in piece-wise linear velocity patterns (roughly triangular in the plots) that indicate clear decision moments when drivers change their control input to help resolve the merging conflict. The model's control behaviour is qualitatively similar to that of the human drivers (Figure~\ref{fig:inputs}-A); it replicates the characteristic patterns in the velocity plots. 

Pair 3 is highlighted in Figure~\ref{fig:inputs}-A to facilitate comparison. In this pair, the left human driver mostly decelerated at the tunnel exit to prevent a collision, although in some cases, they accelerated. The model replicates this behaviour. In one trial (each), the left model and left human maintained their initial velocity for about one second before accelerating. Both the timing of the re-planning (i.e., the location of the peak of the velocity profile) and the absolute maximum deviation from the initial velocity were consistent between the left driver and the left model. Contrary to the left driver, the right driver in pair three barely acted to mitigate the risk. Only in one case (each) the right model and the human driver decelerated at the tunnel exit to prevent a collision. This happened in the same trial where the left driver delayed their initial response and then accelerated for both the model and the human driver pair.
 
The magnitude of control inputs applied by each driver was consistent between the model and human behaviour for most pairs (Figure~\ref{fig:inputs}-B). Some drivers consistently provided very little input (e.g., the left driver in pair 2), while others used higher input levels (e.g., the right driver in pair 7). The model accurately reflected this quantitative difference through the personalised risk thresholds. However, in some specific conditions for specific pairs, the model produced different average behaviour caused by outliers (e.g., pair 1, left driver, condition 0\_8). In this specific example, the simulated left driver came to a complete stop in 2 out of 10 trials. In both cases, the simulated left driver initially accelerated but quickly changed strategies and started to brake. At this point, the right driver had already responded to the initial acceleration and had also started braking. Thus, both vehicles were braking, and the only safe solution was for the left driver to come to a complete standstill. This sequence of events can be understood as a miscommunication caused by a strategy switch. These miscommunications also happened with human drivers (e.g., in Figure~\ref{fig:inputs}-A, the right human driver sometimes decelerates briefly before accelerating and going first). However, with human drivers, these trials ended in a collision or with a less extreme maximum deviation; complete standstills do not occur in the human data. This difference could be due to the velocity perception noise or how communication is translated to a belief in the model. However, since these are edge cases that happen infrequently, a more detailed investigation is needed to understand these miscommunications fully.

Across all participants, both human drivers and the model used lower-magnitude inputs with increasing absolute projected headways (Figure~\ref{fig:inputs}-C). Absolute relative velocities, however, had opposite effects on human and model behaviour. This can partly be explained by the fact that only for human behaviour there is a significant interaction effect. We believe the absence of this effect in the model stems from the independent effects velocity and position have on the belief (i.e., there is no interaction effect in the belief construction). The origins and inner workings of this phenomenon in human behaviour are unknown. In general, there is a strong correlation between the model's and human drivers' inputs (Figure~\ref{fig:inputs}-D) across all conditions and participants. The model exhibited slightly higher maximum absolute deviations from the initial velocity than human drivers (difference $0.31~m/s$, $95~\%$ CI $[0.23~m/s-0.40~m/s]$). We believe this difference is due to the earlier explained outliers.

\begin{figure}[ht!]
    \centering
    \includegraphics[width=\textwidth]{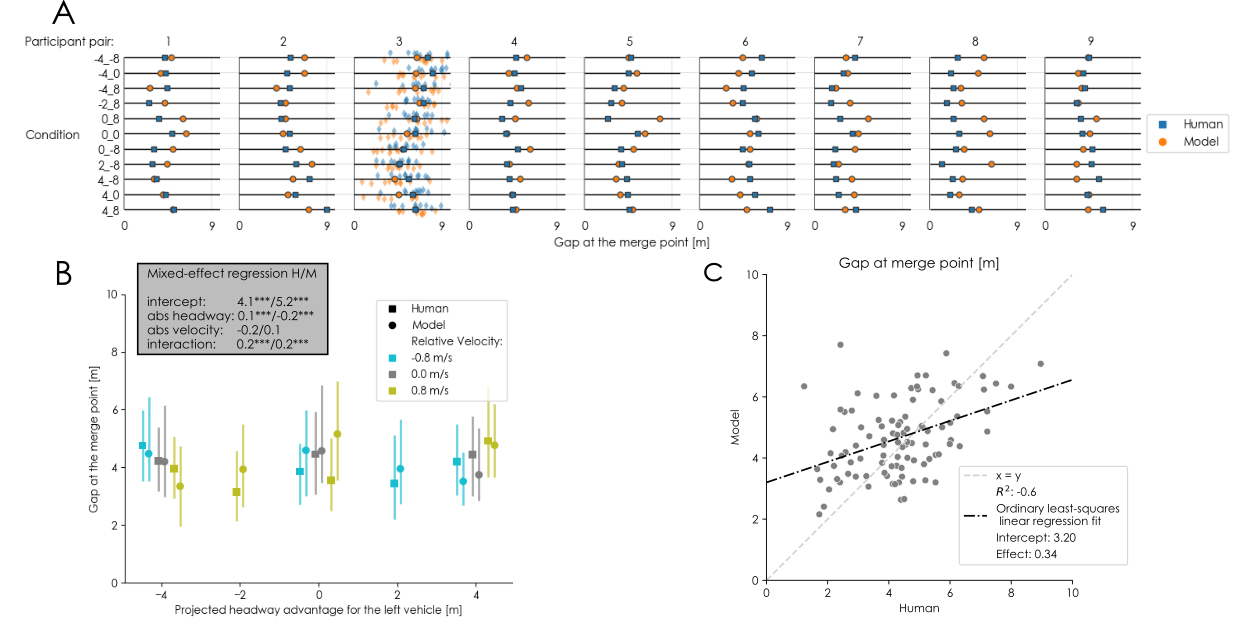}
    \caption{An overview of the joint \textbf{gap-keeping behaviour} of the model and human drivers. \textbf{A}: mean gap (safety margin) at the merge point for all participant pairs and all conditions, excluding collisions. For pair 3, all data points (i.e., all trials) are shown. \textbf{B}: mean gaps per condition (aggregated over all drivers); the error bars represent interquartile ranges. The inset indicates coefficients of mixed-effects linear regression models predicting the mean gap as a function of projected headway and relative velocity ($*: p<=0.05, **: p<=0.01, ***: p<=0.001$); full results of the statistical analysis are available in the supplementary materials. \textbf{C}: the relationship between mean gaps of human drivers and those produced by the model for all participant pairs and all conditions (i.e., all points from panel A).}
    \label{fig:gap}
\end{figure}

\subsubsection*{Safety margin in terms of the size of the gap between the vehicles}
Safety margins can be evaluated individually or at a joint level (Figure~\ref{fig:literature_overview}); drivers individually contribute to achieving a certain realised safety margin (i.e., gap) on the joint level. These individual contributions can be observed from the absolute control input behaviour shown in Figure~\ref{fig:inputs}-B. We use the absolute deviation from the initial velocity to compare conditions because in some conditions drivers accelerate in some trials and decelerate in other trials. 

In participant pair 3, the left driver mostly contributed to the safe solution; the right driver did not greatly deviate from their initial velocity (Figure~\ref{fig:inputs}-B). The model replicates these unequal contributions for this and other pairs (e.g., pairs 2 and 7, and to a lesser extent in pair 5). For the other participant pairs, keeping the safety margin is more of a joint effort, a phenomenon described by the model as well. The model reflects these differences in individual contribution through the baseline risk threshold levels (Table~\ref{tab:intercepts}). Drivers with a lower tolerance for risk (i.e., upper risk threshold) will act to mitigate their perceived risk, while drivers with a higher tolerance will remain passive. Drivers in pairs with equal contributions (e.g., pairs 4, 6, and 9) have similar upper risk thresholds, while drivers in pairs with unequal contributions (e.g., pairs 2, 3, and 7) have larger individual differences (Table~\ref{tab:intercepts}).

In some human pairs, the drivers contribute equally to the safety margin, but their relative contributions differ for different conditions (Figure~\ref{fig:inputs}-B). For example, in pair 6, both drivers always act. However, in the conditions with a projected headway advantage for the left driver (i.e. a positive number), the right driver tends to do more, while for the negative projected headways, the left driver acts. This aligns with the assumption that the following driver has more incentive to solve a conflict. The model captures this behaviour (Figure~\ref{fig:inputs}-B) through the incentive functions. However, in pair 6, the differences between conditions are smaller in the model than in the human behaviour. The limited extent to which the model shows this phenomenon can be attributed to the assumption of identical incentive functions for all pairs. Because only some human pairs show this type of behaviour, the averaged incentive functions reduce the extent to which this phenomenon is visible in the model. A method to estimate individual incentive functions could improve this in future versions of the model.

In terms of joint gap-keeping behaviour, the human data showed no substantial qualitative differences between pairs (Figure~\ref{fig:gap}-A); the effects of kinematic conditions on the gap are significant but small (Figure~\ref{fig:gap}-B). There is much variability within pairs, where gaps range between $0-9~m$ within one condition (pair 3, Figure~\ref{fig:gap}-A) and within conditions, with interquartile ranges of $2-3~m$ (Figure~\ref{fig:gap}-B). This large variability results in a limited correlation between the model and human behaviour, where the model overestimates some smaller gaps and underestimates some larger gaps  (Figure~\ref{fig:gap}-C). However, overall the model keeps gaps that are comparable in size to human behaviour for all participant pairs and all conditions (Figure~\ref{fig:gap}-B). The mean gap for the model was only slightly larger than that of human drivers ($4.8~m$ vs $4.5~m$; difference $0.3~m$, $95~\%$ CI $[0.09~m-0.51~m]$). 

\begin{figure}[ht!]
    \centering
    \includegraphics[width=\textwidth]{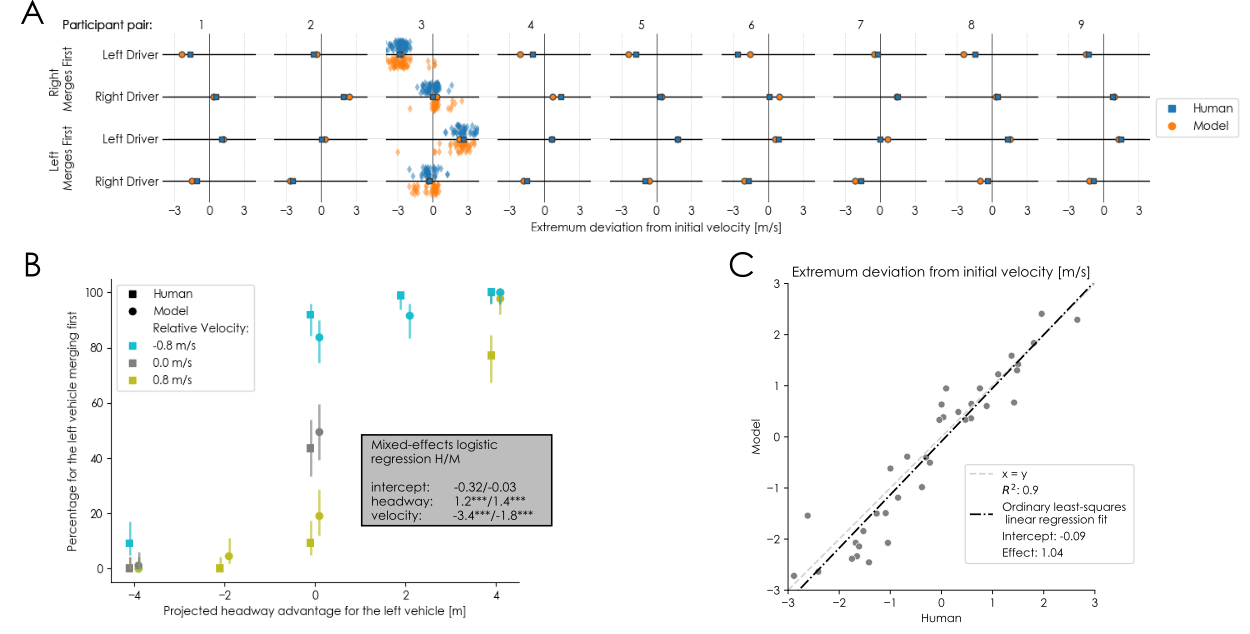}
    \caption{An overview of the \textbf{decision-making behaviour} of the model and human drivers. \textbf{A}: mean individual contributions of drivers to the high-level outcome of a trial as the maximum or minimum deviation from the initial velocity (i.e., amount of acceleration or deceleration); collisions are excluded. All trials are shown for a representative participant pair (pair 3). \textbf{B}: joint interaction outcome (i.e., who merged first) for all pairs in every condition. The error bars represent the $95\%$ confidence intervals. The inset indicates coefficients of mixed-effects linear regression models predicting the mean gap as a function of projected headway and relative velocity ($*: p<=0.05, **: p<=0.01, ***: p<=0.001$); full results of the statistical analysis are available in the supplementary materials. \textbf{C}: the relationship between human and model behaviour for all participant pairs and all conditions (i.e., all points from panel A).}
    \label{fig:who_first}
\end{figure}

\subsubsection*{Decisions of who merges first and who contributes}
\label{sec:decisions}
Finally, there is the high-level outcome of a merging interaction, which on a joint level can be summarised by the answer to the question: "Who merged first?". In human merging interactions, the probability that a driver merges first increases with their projected headway advantage and decreases with their relative velocity advantage~\cite{SiebingaEmpirical}. The model replicates these effects (Figure~\ref{fig:who_first}-B), although the velocity effect is smaller for the model than for human drivers. The difference in this effect size is especially evident in conditions with pure velocity differences (i.e., a $0~m$ projected headway) (Figure~\ref{fig:who_first}-B). In these conditions, the slower vehicle (i.e., the one that merges first most often) approaches the merge point ahead of the other vehicle (so that they arrive simultaneously). Potential explanations for the discrepancy in effect size between the humans and the model could be that humans systematically underestimate velocity differences or that following vehicles prefer braking over accelerating to prevent collisions. The noise on velocity and the cost of accelerating or decelerating are both assumed to be symmetrical in the model. The similar effect sizes indicate that our proposed combination of kinematics-based probabilistic beliefs and the concept of risk-based control in individual drivers is a strong potential explanation of the underlying principles that govern the high-level outcome in merging interactions.

The intercept of the logistic regression (Figure~\ref{fig:who_first}-B) quantifies the asymmetry between outcomes (and participants). The model's intercept is closer to zero than the human drivers', although this effect is insignificant (Figure~\ref{fig:who_first}-B). This can be explained by the fact that the model uses the same symmetrical functions for the belief and incentive across all drivers; therefore, the high-level outcome can be expected to be symmetrical as well (i.e., have an intercept of 0.0). 

The individual contributions of the drivers to the high-level outcomes can be seen as the question for an individual driver to "go or yield"; i.e., the decision to accelerate or to brake (Figure~\ref{fig:who_first}-A). As with the individual contributions to the safety margins, some drivers consistently contribute very little to the high-level outcome (e.g., the left driver in pair 2). The model reflects this phenomenon for multiple drivers using their individual thresholds (left in pairs 2 and 7, right in pairs 3, 5, and 8). However, some drivers only contribute to the high-level outcome when they go second (e.g., the right driver in pair 1), which is reflected by the model through the incentive function. Finally, some drivers always contribute to the outcome (mostly in interactions with drivers that do nothing, e.g., the left driver in pair 3). The model reflects all these three qualitative phenomena (Figure~\ref{fig:who_first}-A). Quantitatively, there is a strong correlation between the humans' and models' decision behaviour (Figure~\ref{fig:who_first}-C).

Besides the individual decisions that lead to a joint high-level outcome, the model also describes how long the drivers take to reach a decision on a safe outcome. This duration can be measured with the Conflict Resolution Time (CRT)~\cite{SiebingaExperiment}. We found that the model captured the previously observed relationship between initial kinematics and CRT (see supplementary materials for details).

\FloatBarrier
\section{Discussion}
We presented a model based on the Communication-Enabled Interaction (CEI) framework~\cite{SiebingaCEITheory} that accurately describes driver behaviour in a simplified interactive merging scenario (Figure~\ref{fig:scenario}-A). Our model captures the driver behaviour on three levels (Figure~\ref{fig:literature_overview}-C): the control input behaviour of individual drivers, the safety margins kept by pairs of drivers and how individual contributions establish these, and the high-level decisions of individual drivers (i.e., to merge or yield) and the pair (i.e., who goes first?). Because the model quantitatively and qualitatively captures individual and joint driver behaviour on all three levels we consider it likely that the underlying mechanisms of the model (communication-based belief and risk-based re-planning) correspond to the mechanisms underlying human interactive driving behaviour. 

These underlying mechanisms bear a resemblance to mechanisms previously used in models of traffic interactions. The communication-based belief is related to the concept of Theory of Mind (ToM)~\cite{Premack1978}, used in other models of traffic interactions (e.g.~\cite{Markkula2023, Tian2021}). The main difference is that ToM assumes humans to have an internal representation of the motivations of others, while our model uses a more basic belief of future kinematics; i.e., our model does not care if another driver prioritises speed over safety, it observes a higher velocity and updates a kinematic belief.   

Risk-based re-planning is a mechanism previously used by Kolekar et al. to model isolated driver behaviour in 7 real-world driving scenarios~\cite{Kolekar2020}. Our model extends the concept of re-planning when the perceived risk exceeds a threshold to an interactive scenario. However, our definition of risk (perceived probability of a collision) is much more simplified than the Driver Risk Field (DRF) used by Kolekar et al. The DRF considers the risk posed by different events (e.g. going off-road or colliding with a tree) and includes steering behaviour. Our simplified scenario does not require such a sophisticated definition. However, including the DRF in a new model based on the CEI framework could enable the modelling of real-world merging scenarios.

Finally, the prevalent approach to modelling multiple drivers in interactions is game theory (e.g.,~\cite{Gabler2017, Camara2020, Nikolaidis2017, Turnwald2014, Tian2021}). Our results have shown that including multiple drivers in a single mode (rather than modelling a single driver that responds to their environment) is important because the same joint behaviour can stem from multiple individual contributions, and both drivers continuously update their behaviour based on the other driver's actions. However, for traffic interactions between vehicles and pedestrians, it has been shown that using game theory to optimise a short-term payoff value is not enough to explain the complex phenomena observed in the real world; instead, a range of more complex mechanisms such as a ToM and implicit communication were needed~\cite{Markkula2023}. 

Our model and the CEI framework can have important further implications on multiple fronts. The model could be used to further improve our understanding of interactive driving behaviour for the development of automated driving technologies, and our model could potentially be generalised to other scenarios with traffic interactions.

The underlying mechanisms of the model enabled it to replicate human driving behaviour on multiple levels. Therefore, the model could help researchers to understand better how these mechanisms function in human behaviour~\cite{Guest2021}. For example, implicit communication (through vehicle movements) and how it influences a driver's belief is observed often but only partly understood~\cite{Brown2023a, Lee2021}. The same holds for the perception of risk, which has been investigated for isolated drivers~\cite{Kolekar2020, Kolekar2020a}, but our model could help extend this to interactions. Our model could facilitate research in these directions, leading to an increase in fundamental knowledge. 

This knowledge could facilitate more practical applications, such as the design of movements that convey a clear message about the intent of an automated vehicle~\cite{Brown2023b}. These movements could consider other drivers' expectations regarding high-level outcomes and communicative actions. Matching the automated vehicle behaviour with expectations might increase behavioural acceptance, although this should be investigated further. Second, the model might be used to inform the real-time interaction planning of automated vehicles. In particular, it could inform the AV about the potential future actions of other drivers (as prediction models are generally being used~\cite{Sadigh2018, Schwarting2019, Ward2017}). Finally, our model could also be valuable in the development phase of automated behaviour by being part of an interactive and dynamic environment for benchmark testing where models are used to evaluate the behaviour of autonomous vehicles (e.g.,~\cite{Queiroz2022, Li2018}). 

Finally, the scenario we modelled in this work bears a resemblance to other interactive scenarios. In essence, the scenario in Figure~\ref{fig:scenario}-A entails a continuous and dynamic interaction where participants search for a mutually beneficial solution. Although there are minor individual advantages to be gained regarding speed and comfort, the most important goal is mutual: to prevent a collision. These aspects are comparable to other interactive scenarios such as traffic interactions between vehicles and pedestrians (e.g.,~\cite{Camara2020, Markkula2023}), pedestrian interactions in a crowd~\cite{Martinez-Gil2017}, or physical human-robot interactive tasks~\cite{DeSantis2008} (e.g.,~\cite{Sadrfaridpour2018, Nikolaidis2017}). The shortcomings of other models that led to the development of the CEI framework~\cite{SiebingaCEITheory} -- the assumption of rationality, absence of communication, and difficulties in extending game theory beyond high-level decisions -- also apply to these related scenarios. Thus, exploring CEI-based models, such as the one presented here, for other interactive scenarios can be an interesting topic for future work.

Our work has three important limitations: the simplified scenario, the manually chosen model parameters, and the complexity of fitting an intermittent behaviour model. To start with the first limitation, we have used a simplified merging scenario in this work to gain insight into the complex dynamics of driver interactions~\cite{SiebingaExperiment}. This scenario enabled us to uncover the characteristics of human behaviour regarding accelerations but does not include two important aspects of merging: steering and traffic rules (such as the right of way). We previously found that the characteristics of human behaviour in our simplified simulator correspond to those found in real-world driving~\cite{SiebingaEmpirical, Markkula2014, Markkula2018}. Therefore, we are confident that our model captures an important aspect of human behaviour in real traffic: intermittent piece-wise constant control. Since the other underlying principles of the model (i.e., risk perception~\cite{Kolekar2020}, communication~\cite{Lee2021, Brown2023a}, and a belief about intent~\cite{Markkula2023}) have also been observed in real traffic and used in other successful models we believe that our model can be generalised to realistic merging scenarios. Nonetheless, extending the model to realistic scenarios (especially realistic beliefs that cover multiple tactical responses to the same situation~\cite{SiebingaHaussdorf}) remains a topic for future work. 

Second, the model uses 10 manually chosen parameters to match the scenario (these were heuristically determined, based on literature, or tuned to fit the data; see the Methods section for details). Among these parameters are the length of the planning horizon for the drivers and the saturation time $\tau$ that governs how long it takes drivers to re-plan when the conflict is resolved. Although we found the model robust to changes in these parameters, how they generalise to other scenarios (e.g., with different dimensions of the track) is unknown. It is possible that all scenarios need a specific set of parameters and that to generalise the model to work in multiple scenarios simultaneously, these parameters need to be dynamically adjusted. 

Finally, due to the intermittent nature of our model, it is complex to fit it to human data. Because of the mechanism where a risk threshold triggers a planning update, individual trials only provide limited information about the threshold value that would describe a driver best. If the driver acts, the threshold is exceeded, and if they don't act, the threshold is not exceeded. The amount of action is not related to the upper risk threshold. Therefore, we used a grid search method to obtain individual values for risk thresholds. However, this method is computationally inefficient, imprecise, and hard to use with more complex scenarios with multiple control inputs (i.e., when including steering). Because intermittent control is a key aspect of the CEI framework and our model, this is a fundamental limitation to the potential to generalise the model. More work is needed to develop a more robust method to fit intermittent models such as ours. 

In conclusion, our model hypothesised that a communication-enabled kinematic belief combined with risk-based intermittent actions underlie human interactive behaviour in merging. In contrast to the currently prevalent game-theoretic models of traffic interactions, our model does not rely on the assumption of rationality and explicitly includes implicit communication between drivers. Despite its simplicity, our model could accurately describe the joint behaviour of human drivers and their individual contributions in merging interactions on three levels: control inputs, safety margins, and decisions. We believe our model could be a useful tool to increase the fundamental understanding of the effects a vehicle's kinematics actions have on the beliefs of other drivers. Therefore, we hope our model represents a step towards understanding driving interactions and developing interaction-aware automated driving.

\section{Methods}
In this work, we evaluate a Communication-Enabled Interaction (CEI) model in a simulated environment and compare it to human behaviour data that was previously collected in a simulator experiment. Here we reiterate the details of the experiment, present the design of the four modules of the CEI model (plan, communication, belief, and risk perception), and discuss the parameters we used for the model and the fitting procedure. Finally, we present the details of the software and data we used in this work which is all available online from public repositories.

\subsection{Experiment and simulation environment}
The data on human driver behaviour we used in this work was previously gathered in an experiment in a coupled top-down-view driving simulator~\cite{SiebingaEmpirical, SiebingaExperiment}. Eighteen volunteers (6 female, 12 male, mean age: 25, std: 2.6) participated in the study and were divided into 9 fixed pairs (i.e., each participant interacted with the same counterpart in all trials). This experiment was approved by TU Delft's Human Research Ethics Committee and all participants gave their informed consent before participating.

The participants controlled the acceleration of their vehicle using the gas and brake pedal of a steering-wheel game controller (Logitech Driving Force GT). The headings of the vehicles were fixed (i.e., equal to the heading of the road). Participants each sat behind a computer screen that showed a top-down view of the simulation. They were seated in the same room behind a curtain to prevent them from seeing each other. The drivers wore noise-cancelling headsets (Sony WH-1000XM3) with ambient music to prevent them from communicating in any other way than through vehicle kinematics. 

In the experiment, the drivers followed a track consisting of three sections of equal length ($50~m$ each, total track length $150~m$): a tunnel, an approach and a car-following section. The vehicle dimensions were $4.5~m$ x $1.8~m$. In the tunnel, the drivers could observe both vehicles and their initial velocity. The initial velocities were either equal ($10~m/s$) for both vehicles or one of the vehicles had a $0.8~m/s$ advantage ($9.6~m/s$ - $10.4~m/s$). If the vehicles maintained their velocity, they would collide at the merge point with varying headways (i.e., distance from front bumper to front bumper). We called this the projected headway and varied it between $0$, $2$, and $4~m$ for each vehicle. Conditions were labelled according to the projected headway and relative velocity (e.g., $-4\_8$) where positive numbers denote an advantage for the left vehicle. Drivers were told: "\textit{Maintain your initial velocity yet prevent a collision. No vehicle has the right of way. Remain seated, use one foot on the gas or brake pedal, keep both hands on the steering wheel, and do not communicate by making sounds or noise. Remember that this is a scientific experiment, not a game or a race.}"

The participants approached the merge point from the left or the right side (randomised before each trial). However, only their own view was varied; in the experimenter's view, and in all results discussed here, the same driver in a pair is referred to as the left or right driver. To facilitate velocity perception, the steering wheels provided vibration feedback when vehicles deviated from their initial velocity. This feedback increased with the magnitude of the deviation. In case of a collision, the simulation was paused for 20 seconds. This time penalty lasted longer than the duration of a single trial ($\sim 16$ seconds), thereby providing an incentive to avoid collisions.

Both in the experiment and the model simulations, we simulated the vehicles as point mass objects, their dimensions were only used for collision detection. The vehicles were subject to a negative acceleration due to resistance of $a_r = 0.5 + 0.005 v^2$, where $v$ is the vehicle's velocity. Vehicle velocities were always positive.

\subsection{Model}
Our proposed model is based on the CEI modelling framework~\cite{SiebingaCEITheory}. According to this framework, joint driver behaviour can be understood as a combination of four modules: plan, communication, belief, and risk perception. 

\paragraph{Plan}
The drivers are assumed to have a deterministic plan for the near future. The model plans to maintain a constant acceleration input over its planning horizon; the acceleration value is obtained by minimising a cost function over this horizon (Equation~\ref{eq:cost}). This cost function $c$ penalises deviating from the desired velocity $v_d$ (which in the experiment is equal to the initial velocity) and large values of the acceleration input $a_t$
\begin{equation}
    c = \sum^{T} (v_t - v_d) ^ 2 + a_t^2,
    \label{eq:cost}
\end{equation}
where $v_t$ denotes the velocity at time $t$, and $T$ is the time horizon. Importantly, the cost function does not include a term for collision avoidance; instead, the CEI framework assumes that drivers manage safety by keeping the risk below the risk threshold, which is imposed as an optimisation constraint on the planning module of the model~\cite{SiebingaCEITheory}. 

With the optimal constant acceleration, a trajectory (i.e., a set of waypoints over time) is constructed over the time horizon $T$. This trajectory is later used to evaluate the perceived risk. The planned constant acceleration is applied with added noise to execute the plan. This noise represents the discrepancy between a planned acceleration and the gas pedal input (i.e., inaccuracies in the driver's neuromuscular system and internal model). The noise is added after the optimisation in the planning phase and remains constant until the next re-plan. Noise is drawn from a scaled normal distribution: $\mathcal{N}(\mu_n=0, \sigma_n^2=\frac{1}{40^2})$.

If the optimisation fails because no solution can be found within the constraints, the model falls back to either full braking or full acceleration. If the ego vehicle is behind the other vehicle, heading for a collision, and no solution to the planning problem can be found, the model applies full braking.  Similarly, if the ego vehicle is ahead of the other vehicle but cannot find a feasible plan, it applies full acceleration. In both cases, a new optimisation is triggered at the next time step until a valid solution can be found again.

\paragraph{Communication}
In our model, drivers communicate through vehicle kinematics. Explicit communication (e.g., with indicator lights) is not included in the model or the experiment for simplicity. Drivers observe the other vehicle's position, velocity, and acceleration at every time step. Position and acceleration observations are assumed to be perfect. Velocity perception is assumed to be noisy to account for the fact that drivers sometimes accelerate and sometimes decelerate at the tunnel exit in the same condition (Figure~\ref{fig:inputs}-A). This behaviour can be explained by the fact that drivers under- or overestimate the other vehicle's velocity in the tunnel.

The noise in the velocity perception is inspired by evidence accumulation, a concept used in driver decision-making studies before~\cite{Zgonnikov2022, Markkula2023}. Specifically, we assume that drivers update their perceived velocity of the other vehicle $v^p$ at every time step with an observation affected by noise
\begin{align}
    v^p_t &= v^p_{t-1} + d v^p \label{eq:vel_p} \\
    d v^p &= \alpha (v_t - v^p_{t-1}) + \beta dW. \label{eq:vel_update}
\end{align}
In Equations~\ref{eq:vel_p} and~\ref{eq:vel_update}, subscript $t$ denotes time, the superscript $p$ denotes the perception of the other vehicle's velocity, $d v^p$ is the perception update, $v$ is the other vehicles true velocity, $\alpha$ denotes the update rate, $\beta$ the noise level, and $W$ is a stochastic Wiener process (thus $dW$ is a sample from a normal distribution $\mathcal{N}(\mu=0, \sigma^2=dt)$).

\paragraph{Belief}
The observed communication is used to create a belief about the future positions of the other vehicle. This belief consists of belief points at a specific belief frequency $f_b$ over the same time horizon $T$ as the one used in the plan. Every belief point is a probability distribution over positions. Each of these belief points are represented by the sum of two normal distributions:

\begin{equation}
    b_t = \frac{1}{2}\mathcal{N}(\mu_t, \sigma^2_t) + \frac{1}{2}\mathcal{N}(\mu_t, \phi \sigma^2_t),
\end{equation}

where $b$ is the belief point representing the probability distribution over positions for the other vehicle at time $t$ and $\phi$ is a scaling factor. The first part of this equation represents the positions of the other vehicle that are kinematically feasible within the bounds of comfortable acceleration. The second part is kinematically infeasible within comfortable bounds and can be interpreted as a belief that something unexpected will happen (e.g., an emergency braking).
The motivation behind including two components to the belief distribution was that a single normal distribution either assigns similar probabilities to the kinematically likely and unlikely positions (when $\sigma_t$ is high) or only considers high-risk scenarios (when $\sigma_t$ is low). Our belief model addresses this issue by emphasising the kinematically likely outcomes, but at the same time including a safety margin in case of errors or unlikely events (e.g., emergency braking, long reaction times, or a perception error of the other driver).

The parameters $\mu_t$ and $\sigma_t$ are based on a normally distributed expected acceleration ($\mathcal{N}(\mu_a, \sigma^2_a)$) which is constructed based on driver's memory about recent acceleration observations. Drivers keep a memory of recent acceleration observations of the other vehicle $M_a$ 
\begin{align}
    M_a &= [a_{-T_m}, .., a_t] \label{eq:memory}\\
    \mu_a &= \bar M_a \label{eq:acc_mean}\\
    \sigma^2_a &= (\frac{1}{3} a_c)^2 + \textrm{var}(M_a)
\end{align}
 
Here, the mean expected acceleration $\mu_a$ is calculated as the average of the remembered values over the past $T_m$ seconds (Equation~\ref{eq:memory}). The standard deviation $\sigma_a$ of the expected acceleration is based on the maximum comfortable acceleration $a_c$ (assuming that $99.7\%$ of observed accelerations fall within $\pm a_c$) and an added variance $\textrm{var}(M_a)$. The latter part increases the expected variance in future accelerations if inconsistent behaviour has recently been observed.

The parameters for the belief point distributions are then constructed using point mass kinematics with this normally distributed acceleration
\begin{align}
    \mu_t &= \frac{1}{2} (t-t_0)^2 \mu_a + v_0 (t-t_0) + p_0, \\
    \sigma^2_t &= \frac{1}{2} (t-t_0)^2 \sigma_a^2,
\end{align}
where $p_0$ denotes the observed position of the other vehicle, $v_0$ is the perceived (noisy) velocity of the other vehicle, and $t$ denotes the time of this belief point and $t_0$ the current time.

\paragraph{Risk perception}
Risk perception combines the planned trajectory and the belief about the future positions of the other vehicle to calculate the probability of a collision. For every belief point, the model determines the bounds of collisions~\cite{SiebingaCEITheory}; these are the extremum positions of the other vehicle that will result in a collision. The believed probability that the other vehicles will be between these bounds is the perceived probability of a collision. This probability is assumed to be the perceived risk. 

The perceived risk is evaluated against two dynamic risk thresholds, the upper ($\rho_u$) and the lower ($\rho_l$) threshold. Both thresholds consist of a driver's individual base value ($\theta$), which is adjusted with an incentive function:
\begin{align}
    \rho_u^d = \theta_u^d + \lambda_{u, 1} \Delta p + \lambda_{u, 2} \Delta v + \lambda_{u, 3} \Delta p \Delta v \label{eq:up}\\
    \rho_l^d = \theta_l^d + \lambda_{l, 1} \Delta p + \lambda_{l, 2} \Delta v + \lambda_{l, 3} \Delta p \Delta v. \label{eq:low}
\end{align}
In these equations, superscript $d$ denotes a specific driver, $\Delta p$ and $\Delta v$ are the relative position and velocity from this driver's perspective, and $\lambda$ are the incentive parameters which are assumed to be constant over the population.

A re-plan is triggered if the upper risk threshold $\rho_u$ is exceeded. This re-plan aims to find an acceleration that brings the perceived risk below $0.8 \rho_l$. If the perceived risk stays below the lower threshold $\rho_l$ for longer than the saturation time $\tau$, the conflict is assumed to be resolved, and another re-plan is triggered to revert to "normal" behaviour. In this case, the risk is constrained to be below $0.6 \rho_u$. Finally, if the desired velocity is reached while the vehicle is accelerating or decelerating, another re-plan is triggered to allow the vehicle to maintain the preferred velocity. 

\subsection{Model parameter fitting}
\label{sec:fitting}
Our model and simulation use 10 parameters with values that were manually designed, their values are shown in Table~\ref{tab:constants}. The timing parameters for the simulation ($dt$), planning ($T$), and belief ($T_m, f_b$) were chosen such that they are suitable for the scenario yet enable reasonable computation times. The noise parameters $\sigma_n$ and $\beta$, and saturation time $\tau$ were manually tuned to reflect the human data. The belief scaling factor $\phi$ was designed to obtain sufficient resolution in the risk signals. The parameters $\alpha$ (which denotes how long drivers need to observe a vehicle to estimate its velocity) and $a_c$ (the maximal comfortable acceleration) were based on empirical literature: $\alpha$~\cite{Fath2018, Jorges2022}, $a_c$~\cite{Hoberock1977}.

\begin{table}[h!]
    \caption{Model and simulation parameters with constant values for all participants}
    \begin{subtable}[t]{.45\textwidth}
    \centering
    \caption{Manually designed parameters}
    \label{tab:constants}
    \begin{tabular}{|ll|ll|} \hline
        $T$ & $6.0~s$ & $dt$ & $0.05~s$  \\
        $T_m$ & $4.0~s$ &  $f_b$ & $4~Hz$ \\ 
        $\sigma_n$ & $\frac{1}{40}$ &  $\beta$ & $0.6$\\
        $\tau$ & $1.6~s$ & $\phi$ & $3.0$\\
        $\alpha$ & $0.5$  &  $a_c$ & $1.0~\frac{m}{s^2}$\\ \hline
    \end{tabular}
    \end{subtable}
\begin{subtable}[t]{.5\textwidth}
    \centering
    \caption{The fitted population-level parameters for the incentive functions}
    \label{tab:incentive}
    \begin{tabular}{|c|c|c|c|c|}\hline
        Parameter & Value & Std. Err. &  z  & p \\ \hline
        $\lambda_{u,1}$ & 0.003 & 0.001 & 3.27 & 0.001 \\
        $\lambda_{u,2}$ & 0.018 & 0.006 & 2.97 & 0.003 \\
        $\lambda_{u,3}$ & -0.006 & 0.001 & -6.38 & 0.000 \\
        $\lambda_{l,1}$ & 0.004 & 0.001 & 3.55 & 0.000 \\
        $\lambda_{l,2}$ & 0.016 & 0.008 & 1.95 & 0.051 \\
        $\lambda_{l,3}$ & -0.003 & 0.001 & -2.03 & 0.042 \\ \hline
    \end{tabular}
    \end{subtable}
\end{table}

The risk thresholds and incentive function parameters were fitted to the data using a grid search. We created a $25 \times 25$ grid for every kinematic condition using upper thresholds in the range $[0.3, 0.9]$ and lower thresholds in the range $[0.01, 0.4]$. For these grids, we disabled the incentive functions and only used the base thresholds ($\theta$ in Equations~\ref{eq:up} and~\ref{eq:low}). We ran one simulated trial per set of thresholds per condition with all the noise in the model disabled, resulting in 11 grids of 625 trials.

In every trial, we simulated the behaviour of a single CEI driver against a vehicle travelling at constant velocity to obtain the immediate response of the (CEI) driver at the tunnel exit before any interaction takes place. For every such trial, we recorded the modelled driver's deviation from the initial velocity after $1.0$ second.

Then, for every trial of every human participant, we searched the grid for the risk threshold values that best described that driver's velocity deviation after $1.0$ second. This resulted in individual base threshold values ($\rho_u$ and $\rho_l$) for every trial, in total $110$ sets of thresholds for every participant. With 18 participants, this gave us 1980 upper and lower threshold values for 11 different kinematic conditions.

To determine participant-level thresholds based on these trial-level thresholds, we used two linear mixed-effect models, one for the lower and one for the upper threshold. Both models used the form: $\rho \sim \Delta p * \Delta v$, where $\rho$ denotes the threshold and $\Delta p$ and $\Delta v$ are the relative position and relative velocity respectively. Random intercepts were included per participant. The resulting fixed-effect coefficients (Table~\ref{tab:incentive}) were used as incentive parameters ($lambda$ in Equations~\ref{eq:up} and ~\ref{eq:low}), and the random-effect coefficients (Table~\ref{tab:intercepts}) were used as participant-level base values $\theta_l$ and $\theta_u$.

\begin{table}[h]
    \centering
    \caption{Personal base values for the upper and lower thresholds for each driver}
    \label{tab:intercepts}
    \begin{tabular}{|c|l|c|c|}\hline
        Pair&Driver& $\theta_l$  &$\theta_u$ \\ \hline
        1&left& 0.165 &0.495 \\
        &right& 0.260 &0.562 \\ \hline
        2&left& 0.245 &0.635 \\
        &right& 0.058 &0.493 \\ \hline
        3&left& 0.058 &0.488 \\
        &right& 0.245 &0.631 \\ \hline
        4&left& 0.183 &0.537 \\
        &right& 0.201 &0.524 \\ \hline
        5&left& 0.113 &0.498 \\
        &right& 0.269 &0.585 \\ \hline
        6&left& 0.246 &0.550 \\
        &right& 0.161 &0.546 \\ \hline
        7&left& 0.320 &0.736 \\
        &right& 0.201 &0.522 \\ \hline
        8&left& 0.165 &0.525 \\
        &right& 0.246 &0.586 \\ \hline
        9&left& 0.178 &0.519 \\
        &right& 0.227 &0.543 \\ \hline
    \end{tabular}
\end{table}

\subsection{Software and data}
We implemented a custom simulation environment in Python to run the experiment and simulate the model. The optimisation in the planning part of the model was implemented with Casadi~\cite{Andersson2018}. The statistical analyses were performed in python using the statsmodels package~\cite{seabold2010statsmodels}.

Our code is publicly available on Github~\cite{SiebingaMergingRepo}. The data from the experiment~\cite{SiebingaDataExperiment} and model simulations~\cite{SiebingaDataModel} are available on the 4TU data repository. Interactive plots of the human data are available online~\cite{SiebingaEmpiricalSupplement}. 

\section*{Acknowledgement}
This research was partly funded by Nissan Motor Company, Ltd. and by the RVO grant TKI2012P01. We thank Lorenzo for his help with the Casadi implementation and Tom for his valuable feedback on the manuscript.

\clearpage
\bibliographystyle{ieeetr}
\bibliography{my_collection}

\end{document}